\documentclass[reqno,12pt]{amsproc}
\newtheorem{theo}{Theorem}
\newtheorem{rem}{Remark}

\newtheorem{lem}{Lemma}

\newtheorem{df}{Definition}
\newtheorem{hyp}{Hypothesis}

\newcommand\eps\varepsilon
\newcommand\ph\varphi
\newcommand\kap\Lambda

\usepackage{enumerate}
\usepackage{graphicx}
\usepackage{float}
\usepackage{tikz-cd}

\begin{document}

\title[Infinite Dimensional  Lagrange-D'Alembert Principle ]
{On Infinite Dimensional Version of the Lagrange-D'Alembert Principle }

\author[Oleg Zubelevich]{Oleg Zubelevich\\ \\\tt
 Steklov Mathematical Institute of Russian Academy of Sciences\\
Mechanics and Mathematics Faculty,\\
M. V. Lomonosov Moscow State University\\
Russia, 119899, Moscow,  MGU \\ozubel@yandex.ru
 }
\email{ozubel@yandex.ru}
\date{}
\thanks{The research was funded by a grant from the Russian Science
Foundation (Project No. 19-71-30012)}

\begin{abstract}The Lagrange-D'Alembert Principle is one of the fundamental 
tools of classical mechanics. We generalize this principle to mechanics-like ODE in Banach spaces. 

 As an application we discuss geodesics in infinite dimensional manifolds and a random ODE with nonholonomic constraint.
\end{abstract}

\maketitle
\numberwithin{equation}{section}
\newtheorem{theorem}{Theorem}[section]
\newtheorem{lemma}[theorem]{Lemma}
\newtheorem{definition}{Definition}[section]

\section{The  Main Theorems. Discussion}
In this short note we extend the Lagrange-D'Alembert formalism to an infinite dimensional ODE of classical mechanics type. Particularly we develop an infinite dimensional theory of ideal holonomic and nonholonomic constraints.

Other infinite dimensional approach based on the Hamel equations is contained in \cite{Shi}.

We hope that our generalization of the Lagrange-D'Alembert formalism will be of some use as a class of instances for infinite dimensional dynamical systems.

Let $X,Y$ be Banach spaces.

Let $I=(t_1,t_2)\subset \mathbb{R}$ stand for the fixed open interval. Sets $D,\tilde D\subset X$ are open; $M=I\times D\times \tilde D$.

We use $\mathcal B(X',X)$ to denote the space of bounded linear operators $u:X'\to X$. 
Here and below the symbol prime denotes duality.

Being equipped with the norm
$$\|u\|_{\mathcal B(X',X)}=\sup\{\|u(x)\|_{X}\mid \|x\|_{X'}\le 1\}$$ the space $\mathcal B(X',X)$ becomes a Banach space.

Introduce the following functions
$$f\in C^1(M,X'),\quad P\in C^1(M,\mathcal B(X',X)).$$
This means that the function $f:M\to X'$ is  continuously differentiable in $M$ in the sense of  Fr\'echet and for the function $P$ the corresponding assumption is true.

This condition can slightly be  relaxed but to simplify the narration  we do not do that. 

\begin{hyp}\label{2}Assume also that for each $v\in X',\quad z\in M$
it follows that
$$|(v,P(z)v)|\ge K(z)\|v\|_{X'}^2.$$
Here $K$ is a positive valued function.
\end{hyp}
\begin{rem}By lemma \ref{xdg4400} (see below) this hypothesis implies that for each $z\in M$ the operator $P(z):X'\to X$ is an isomorphism. \end{rem}

The main object of our study is the following problem
\begin{equation}\label{xfg4000}
\ddot x=P(z)\big(f(z)+N(z)\big),\quad z=(t,x,\dot x)\in M.\end{equation}
\begin{rem}In section \ref{sdfff} we show that the Lagrange equations have shape (\ref{xfg4000}).\end{rem}
Here the function $f$ is given and a function $N$ must be determined such that $N\in C^1(M,X')$ and  equation of constraints
\begin{equation}\label{xfd11}
\ph(z)=0\end{equation}
determines a manifold $S$ that is invariant under the time shifts of system (\ref{xfg4000}). Here the function $\ph$ belongs to $C^2(M,Y)$.
We surely assume that the set $$S=\{z\in M\mid \ph(z)=0\}$$ is non void.

\begin{hyp}\label{1}For any $z\in M$ the derivative 
$$\ph_{\dot x}(z):X\to Y,\quad \ph_{\dot x}\in C^1(M,\mathcal B(X,Y))$$ is a mapping onto. 
\end{hyp}
If  the function $N\in C^1(M,X')$ then system (\ref{xfg4000}) satisfies all the conditions of the existence and uniqueness theorem \cite{ded}.

In classical mechanics the function $\ph$ is as a rule linear in $\dot x$.

\begin{theo}\label{xfg5500}
Under assumptions above there exists a unique function $N\in C^1(M,X')$ such that

1) the function $\ph$ is a first integral of (\ref{xfg4000})

and

2) for all $z\in M$ the following inclusion holds
\begin{equation}\label{xfbaa}
\ker \ph_{\dot x}(z)\subset\ker N(z).\end{equation}
\end{theo}
\begin{theo}\label{sdg112ww}
An operator $$b(z)=\ph_{\dot x}(z)P(z)\ph_{\dot x}'(z): Y'\to Y$$ is an isomorphism and
$$ b\in C^1(M,\mathcal B(Y',Y)),\quad b^{-1}\in C^1(M,\mathcal B(Y,Y')).$$
The reaction $N$ given by theorem \ref{xfg5500} is expressed as follows
\begin{equation}\label{srsg500oo}
N(z)=-\ph_{\dot x}'(z)b^{-1}(z)\big(\ph_t(z)+\ph_x(z)\dot x+\ph_{\dot x}(z)P(z)f(z)\big).\end{equation}
\end{theo}
\begin{rem}Actually the  restriction $N\mid_S$ does not depend on which a function $\ph$ determines the manifold $S$. We discuss it in details in  section \ref{dfg55ff}.\end{rem}
We prove theorems \ref{xfg5500}, \ref{sdg112ww} in section \ref{12ww}.

In the sequel we assume that the function $N$ is chosen in accordance with theorem \ref{xfg5500}.

The space $\ker \ph_{\dot x}(z)$ is referred to as the space of virtual displacements.

Under conditions of theorem \ref{xfg5500} equation (\ref{xfd11}) is called the equation of ideal constraints and the function $N$ is called the reaction of ideal constraints.

Formula (\ref{xfbaa}) means that the work done by the reaction of ideal constraint $N$ at a virtual displacement $\xi,\quad \ph_{\dot x}\xi=0$ vanishes:
$N\xi=0.$

Condition 1) of theorem \ref{xfg5500} implies in particular that the manifold $S$
is invariant under the phase flow of system (\ref{xfg4000}) that is
if $x(t)\in C^2(I,X)$ is a solution to (\ref{xfg4000}) and for some $t_0\in I$ one has
$$(t_0,x(t_0),\dot x(t_0))\in S$$ then
$$(t,x(t),\dot x(t))\in S\quad \forall t\in I.$$

The following theorem is a direct consequence of formulas (\ref{xfg4000}), (\ref{xfbaa}).

\begin{theo}\label{xdf440}
Let $x\in C^2(I,X)$ be a solution to (\ref{xfg4000}). Then for all $t\in I$ the following inclusion holds
\begin{equation}\label{dfh09}
\ker \ph_{\dot x}\big(t,x(t),\dot x(t)\big)\subset\ker\Big(P^{-1}\big(t,x(t),\dot x(t)\big)\ddot x(t)-f\big(t,x(t),\dot x(t)\big)\Big).\end{equation}\end{theo}
Formula (\ref{dfh09}) is called the Lagrange-D'Alembert equation. The Lagrange-D'Alembert equation is free from the reaction of ideal constraints $N$.

The converse assertion holds.

\begin{theo}\label{xfg009dd}
Assume that a function $x\in C^2(I,X)$ satisfies (\ref{xfd11}) and (\ref{dfh09}). Then $x$ satisfies (\ref{xfg4000}).\end{theo}
Theorem \ref{xfg009dd} is proved in section \ref{xcb00}.

\subsection{The  Lagrange Equations}\label{sdfff}
Consider a Lagrangian
$$L(z)=T(z)-V(t,x),\quad T=\frac{1}{2}\big(\dot x,G(t,x)\dot x\big)$$
where $$V\in C^2(I\times D,\mathbb{R}),\quad G\in C^2\big(I\times D,\mathcal B(X,X')\big).$$
Assume that the following condition holds:
$$\big(\xi,G(t,x)\eta\big)=\big(G(t,x)\xi,\eta\big),\quad \xi,\eta\in X$$
and thus $T_{\dot x\dot x}=G$.

The corresponding Lagrange equations of the second kind are
\begin{equation}\label{dfg600kk}\frac{d}{dt}L_{\dot x}-L_x=Q+N,\end{equation}
where $Q\in C(M,X')$ is a given generalized force.

Equation (\ref{dfg600kk}) has  shape (\ref{xfg4000}) with
$$f=Q(z)-T_{\dot x t}(z)-T_{\dot x x}(z)\dot x-V_x(t,x)+T_x(z)$$
provided $P=G^{-1}$ is well defined and satisfies the conditions above.

\subsection{Holonomic Constraints}\begin{df}
We shall say that constraint (\ref{xfd11}) is holonomic iff there exists a function
$$g\in C^3(I\times D,Y),\quad g= g(t,x)$$ such that the manifold $S$ can equivalently be determined  as follows
$$S=\{z=(t,x,\dot x)\in M\mid g_t(t,x)+g_x(t,x)\dot x=0\}.$$
The mapping $g_x(t,x):X\to Y$ is onto for each $(t,x)\in I\times D$.\end{df}
The function $g$ is evidently defined up to an additive constant function $(t,x)\mapsto \mathrm{const}\in Y$.

The problem whether a constraint is holonomic or not is solved by means of the infinite dimensional
version of the Frobenius theorem \cite{ded}. This situation is analogous to the finite dimensional one.

\begin{theo}\label{xdsg00}
A manifold
$$\Sigma=\{z\in S\mid g(t,x)=0\}$$  is an invariant manifold of system (\ref{xfg4000}).\end{theo}
Indeed, let $x(t)$ be a solution to (\ref{xfg4000}) such that for some $t_0\in I$ one has 
$$\big(t_0,x(t_0),\dot x(t_0)\big)\in \Sigma.$$
The manifold $S$ is invariant:
$$\frac{d}{dt}g\big(t,x(t)\big)=g_x\big(t,x(t)\big)\dot x(t)+g_t\big(t,x(t)\big)=0\quad \forall t\in I.$$
Integrating this equality from $t_0$ to $t$ we obtain $g\big(t,x(t)\big)=0$.

This proves the theorem.

\section{Geodesics on an Infinite Dimensional Manifold}Assume that constraint (\ref{xfd11}) is holonomic and has the following form
\begin{equation}\label{xfg5ll}\ph(t,x,\dot x)=g_x(x)\dot x=0,\quad \tilde D=X.\end{equation}
The function $g\in C^3(D,Y)$ is such that the operator $g_x(x):X\to Y$ is onto for each $x\in D$.

Introduce a manifold 
$$\Gamma=\{x\in D\mid g(x)=0\}.$$

Let the operator $P$ be independent on $z$. 
\begin{df} A curve $\gamma=\{x(t)\in D\mid t\in I\}$ is said to be a geodesic in $\Gamma$ if the function
$x\in C^2(I,X)$ is a solution to the following initial value problem:
\begin{align}\ddot x&=-Pg'_x(x)\big(g_x(x)Pg_x'(x)\big)^{-1}g_{xx}(x)[\dot x,\dot x],\label{dfg55}\\
(t_0,x(t_0),\dot x(t_0))\in\Sigma&=\{(t,x,\dot x)\in M\mid g(x)=0,\quad g_x(x)\dot x=0\}.\label{xdfgert}\end{align}
Here $g_{xx}(x)[\cdot,\cdot]\in \mathcal B\big(X,\mathcal B(X,Y)\big).$
\end{df}
Equation (\ref{dfg55}) corresponds to  equation (\ref{xfg4000}) with constraint (\ref{xfg5ll}) and $f=0$.

By theorem \ref{xdsg00} initial condition (\ref{xdfgert}) guarantees that $\gamma\subset \Gamma$.
\begin{rem}
If the spaces $X,Y$ are finite dimensional and the operator $P$ 
is symmetric: $P'=P$ then $P^{-1}$ is a Riemann metric in $X$. This metric endows the manifold $\Gamma \subset X$ with a Riemann metric in usual way.  

Equation (\ref{dfg55}) describes dynamics of a particle that slides along the surface $\Gamma$ freely. Trajectories of such a particle are the geodesics in $\Gamma$.
\end{rem}

\subsection{Geodesics on Quadric in Hilbert Space}
 Let $X=\ell_2$. Recall that $\ell_2$ is a Hilbert space with respect to an inner product
$$x=(x_1,x_2,\ldots),\quad y=(y_1,\ldots),\quad (x,y)=\sum_{i=1}^\infty x_iy_i.$$
We consider a real version of $\ell_2: x_i,y_i\in\mathbb{R}$. By the Riesz representation theorem this space is identified with its dual: $\ell_2=\ell'_2.$

Let us put 
$$ P=\mathrm{id}_X,\quad g(x)=(x,Wx)-1,$$
where the operator $W\in\mathcal B(\ell_2, \ell_2),\quad W'=W\ne 0$. So that $\Gamma$ is a quadric,
$$D=\ell_2\backslash \ker W,\quad \tilde D=X.$$

Then the equation of geodesics (\ref{dfg55}) takes the form
\begin{equation}\label{sxdfg5005}\ddot x=-\frac{(\dot x,W\dot x)}{\|Wx\|_{\ell_2}^2}Wx.\end{equation}
If $$\{e^{\Omega s}\mid s\in\mathbb{R}\},\quad \Omega\in\mathcal B(\ell_2,\ell_2),\quad \Omega'=-\Omega$$
is a symmetry group of the quadric $\Gamma$:
$$(e^{\Omega s}x,We^{\Omega s}x)=1,\quad \forall s\in\mathbb{R}$$ then
$$W\Omega-\Omega W=0$$ and a function $(x,\Omega\dot x)$ is a first integral of (\ref{sxdfg5005}).

It is not hard to show that if $x(t)$ is a solution to (\ref{sxdfg5005}) and $g(x(t))=1,\quad t\in I$ then the
"kinetic energy" is conserved:
$$T=\frac{1}{2}\| x(t)\|^2_{\ell_2}=\mathrm{const}.$$
\section{An Example of Nonholonomic Constraint. Random ODE }
Introduce a probability measure $\mu$ in $\mathbb{R}$ as follows
$$d\mu=\rho(\omega)d\omega,$$ where the function $$\rho\in L^\infty(\mathbb{R})\cap L^1(\mathbb{R})$$ is such that for almost all $\omega\in\mathbb{R}$ it follows that $\rho(\omega)\ge 0$  and
$$\int_{\mathbb{R}}\rho(\omega)d\omega=1.$$

Let us
 put
$$X=L^2(\mathbb{R},\mu),\quad P=\mathrm{id}_X,\quad  D=\tilde D=X\backslash\{0\},\quad f=0.$$
Here we take into account the isomorphism between $X$ and $X'$ and identify $X=X'$.

A function $$x=x(t,\omega),\quad x:I\to X$$ is called the second order stochastic process \cite{eimling}.

Introduce the constraint as follows
\begin{equation}\label{srg500}\ph(t,x,\dot x)=\frac{1}{2}\Big(\|\dot x\|^2_{X}+\|x\|^2_{X}\Big)-1=0.\end{equation}
This function can be interpreted as the full energy of the continuum of independent linear oscillators with the same frequencies. The  constraint
makes the system to move such that the energy is conserved. Let us show that the system behaves like an oscillator indeed.

Equation (\ref{xfg4000}) takes the form
\begin{equation}\label{xvf55}\ddot x=-\frac{(x,\dot x)}{\|\dot x\|^2_{X}}\dot x.\end{equation}Its solution is presented as follows
\begin{equation}\label{drfhhgqq} x(t,\omega)=C_1(\omega)+u(t)C_2(\omega),\quad C_1,C_2\in X,\quad C_2\ne 0,\end{equation}
 where the scalar function $u$ satisfies the following IVP
$$\ddot u+u+\frac{(C_1,C_2)}{\|C_2\|_{X}^2}=0,\quad u(0)=0,\quad \dot u(0)=1;$$
and
$$C_1(\omega)=x(0,\omega),\quad C_2(\omega)=\dot x(0,\omega).$$
The last formulas can be interpreted such that we have a unique oscillator with distributed initial conditions.

The classical solution to  problem (\ref{xvf55}) exists until $\dot u= 0$; but we can treat formula (\ref{drfhhgqq})  as a generalized solution defined for all real $t$.

\section{Proof of Theorems \ref{xfg5500}, \ref{sdg112ww}}\label{12ww}
We need the following facts.
\begin{theo}[\cite{KF}]\label{xfg77}
Let $E,F,H$ be Banach spaces. And operators
$$A\in \mathcal B(E,H),\quad B\in \mathcal B(E,F)$$ be such that $B$ is onto and $\ker B\subset\ker A$.

Then there exists an operator $\mathcal L\in \mathcal B(F,H)$ such that  the following diagram

\begin{tikzcd}
& F \arrow[dd, dashed, "\mathcal L"] \\
E \arrow[ur,"B"] \arrow[dr, "A"] & \\
& H
\end{tikzcd}

is commutative: $A=\mathcal L B$.\end{theo}

\begin{theo} [\cite{edvards}]\label{sdf22}An operator $\psi\in \mathcal B(E,F)$ is onto iff  the operator $\psi':F'\to E'$ is a strong isomorphism between $F'$ and $\psi'(F')$. 

The operator $\psi'\in \mathcal B(F',E')$ is onto iff  the operator $\psi:E\to F$ is an isomorphism between $E$ and $\psi(E)$. 
\end{theo}
In the sequel  $c_1,c_2,\ldots$ denote inessential positive constants.
\begin{lem}\label{xdg4400}Assume that an operator $J\in\mathcal B(E',E)$ is such that
\begin{equation}\label{400}|(w,Jw)|\ge c_1\|w\|_{E'}^2,\quad \forall w\in E'.\end{equation}
Then the operator $J$ is an isomorphism between $E'$ and $E$.\end{lem}
{\it Proof of lemma \ref{xdg4400}.}
It is clear that the operator $J$ is an injection. Take any $\xi\in J(E')$. Then we have
$$c_1\|J^{-1}\xi\|_{E'}^2\le |(J^{-1}\xi,\xi)|\le \|J^{-1}\xi\|_{E'}\cdot \|\xi\|_{E}.$$
Thus $J$ is an isomorphism between $E'$ and $J(E')$. By theorem \ref{sdf22} the operator $$J':E'\to E''$$ is onto.

The operator $J'$ is one-to-one. This follows by formula $(w,Jw)=(J'w,w)$ from inequality (\ref{400}). 

From the Open Mapping theorem we conclude that $J'$ is an isomorphism.

By theorem \ref{sdf22} the operator $J$ is an isomorphism between $E'$ and $E$.

Lemma \ref{xdg4400} is proved.

\begin{theo}\label{u5u5}Let  $E,F$ be Banach spaces; $$A\in \mathcal B(E',E),\quad B\in \mathcal B(E,F).$$
 The operator
  $B$ is onto. The operator $A$ satisfies the following inequality
  $$|(w,Aw)|\ge c_2\|w\|^2_{E'},\quad \forall w\in E'.$$

  Then the operator
  $$R=BAB':F'\to F$$ is an isomorphism.\end{theo}
{\it Proof of Theorem \ref{u5u5}.}
From  theorem \ref{sdf22} it follows that $B':F'\to B'(F')\subset E'$ is an  isomorphism. We consequently have 
$$\|B'u\|_{E'}\ge c_3\|u\|_{F'}.$$
It follows that
\begin{equation}\label{xdfbj}|(w,Rw)|=|(B'w,AB'w)|\ge c_2\|B'w\|^2_{E'}\ge c_4\|w\|_{F'}^2.\end{equation}
Now the assertion follows from lemma \ref{xdg4400}.

Theorem \ref{u5u5} is proved.

We are ready to prove theorems \ref{xfg5500}, \ref{sdg112ww}.

First suppose that the reaction  $N$ exists and prove its uniqueness. 

From formula (\ref{xfbaa}) and theorem \ref{xfg77} it follows that for each $z\in M$ there exists an operator 
$\Lambda(z)\in Y' $ such that
\begin{equation}\label{2eerr}N(z)=\Lambda(z)\ph_{\dot x}(z)=\ph_{\dot x}'(z)\Lambda(z).\end{equation}
Condition 1) of theorem \ref{xfg5500}
is equivalent to the following equation
\begin{equation}\label{33eee}
\ph_t(z)+\ph_x(z)\dot x+\ph_{\dot x}(z)P(z)\big(f(z)+N(z)\big)=0.\end{equation}
Substituting (\ref {2eerr}) to (\ref{33eee}) and by using theorem \ref{u5u5} we obtain
$$
\Lambda(z)=-b^{-1}(z)\big(\ph_t(z)+\ph_x(z)\dot x+\ph_{\dot x}(z)P(z)f(z)\big).$$

Since $b\in C^1(M,\mathcal B(Y',Y))$ it follows that $b^{-1}\in C^1(M,\mathcal B(Y,Y'))$ \cite{ded}. Therefore we yield (\ref{srsg500oo}) and $N\in C^1(M,X').$

Formula (\ref{srsg500oo}) proves the existence as well.

Theorems  \ref{xfg5500}, \ref{sdg112ww} are proved.

\section{Proof of Theorem \ref{xfg009dd}}\label{xcb00}

By theorem \ref{xfg77} inclusion (\ref{dfh09}) implies that there exists a function
$$\tilde \Lambda:I\to Y'$$ such that 
$$
P^{-1}\big(t,x(t),\dot x(t)\big)\ddot x(t)-f\big(t,x(t),\dot x(t)\big)=\ph_{\dot x}'(t,x(t),\dot x(t)\big)\tilde \Lambda(t)$$
or
\begin{equation}\label{x400}
\ddot x(t)=P\big(t,x(t),\dot x(t)\big)\Big(f\big(t,x(t),\dot x(t)\big)+\ph_{\dot x}'\big(t,x(t),\dot x(t)\big)\tilde \Lambda(t)\Big).\end{equation}
Differentiate the equality 
$$\ph\big(t,x(t),\dot x(t)\big)=0$$ to have
$$\ph_t\big(t,x(t),\dot x(t)\big)+\ph_x\big(t,x(t),\dot x(t)\big)\dot x(t)+\ph_{\dot x}\big(t,x(t),\dot x(t)\big)\ddot x(t)=0.$$
Substituting here $\ddot x$ from (\ref {x400}) we make sure that
$$\tilde\Lambda(t)=\Lambda\big(t,x(t),\dot x(t)\big).$$
This completes the proof.

\section{Independence $N\mid_S$ on $\ph$}\label{dfg55ff}
Formula (\ref{srsg500oo}) looks like $N\mid_S$ depends on $\ph.$ Actually it is not so.

Indeed, consider a function 
$$U\in C^1(M\times Y, Y),\quad U=U(t,x,\dot x,y)$$ such that for all $z\in M$
the mapping 
$$U_y(z,0):Y\to Y$$ is an isomorphism and
$$U(t,x,\dot x,y)=0\Leftrightarrow y=0.$$
This particularly implies
\begin{equation}\label{xdf55tt}
U_t(z,0)=0,\quad U_x(z,0)=0,\quad U_{\dot x}(z,0)=0.\end{equation}

Equation (\ref{xfd11}) is equivalent to the following one
\begin{equation}\label{xfg600i}\sigma(z)=0,\end{equation} where 
$$\sigma(z)=U\big(z,\ph(z)\big).$$
In other words equation (\ref{xfg600i}) determines the same manifold $S$. 

Then  one yields
\begin{align}
N\mid_{z\in S}&=-\ph_{\dot x}'(z)b^{-1}(z)\big(\ph_t(z)+\ph_x(z)\dot x+\ph_{\dot x}(z)P(z)f(z)\big)\mid_{z\in S}\nonumber\\
&=-\sigma_{\dot x}'(z)\big(\sigma_{\dot x}(z)P(z)\sigma_{\dot x}'(z)\big)^{-1}\nonumber\\
&\big(\sigma_t(z)+\sigma_x(z)\dot x+\sigma_{\dot x}(z)P(z)f(z)\big)\mid_{z\in S}.\label{dfg00}
\end{align}
Indeed, due to (\ref{xdf55tt}) one has:
\begin{align}
\sigma_t\mid_{z\in S}&=(U_t+U_y\ph_t)\mid_{z\in S}=U_y\ph_t,\nonumber\\
\sigma_x\mid_{z\in S}&=(U_x+U_y\ph_x)\mid_{z\in S}=U_y\ph_x,\nonumber\\
\sigma_{\dot x}\mid_{z\in S}&=(U_{\dot x}+U_y\ph_{\dot x})\mid_{z\in S}=U_y\ph_{\dot x}\nonumber.\end{align}
Now formula (\ref{dfg00}) follows from direct calculation.

By the same reason
$$\ker\big(\sigma_{\dot x}\mid_{z\in S}\big)=\ker\big(\ph_{\dot x}\mid_{z\in S}\big).$$


\end{document}